\documentclass[aps,prl,superscriptaddress,showpacs,floatfix,twocolumn]{revtex4}

\usepackage{graphicx}	% Include figure files

\def\auau{Au+Au }
\def\pp{p+p }
\def\dau{d+Au }
\def\sqrts{$\sqrt{s_{NN}}$}
\def\pt{$p_T$ }
\def\gevc{GeV/$c$ }
\def\raa{$R_{AA}$ }

\begin{document}

\title{Nuclear Modification of Electron Spectra and Implications for
Heavy Quark Energy Loss in \auau Collisions at \sqrts~=~200~GeV}

\newcommand{\abilene}{Abilene Christian University, Abilene, TX 79699, U.S.}
\newcommand{\acadsin}{Institute of Physics, Academia Sinica, Taipei 11529, Taiwan}
\newcommand{\banaras}{Department of Physics, Banaras Hindu University, Varanasi 221005, India}
\newcommand{\barc}{Bhabha Atomic Research Centre, Bombay 400 085, India}
\newcommand{\bnl}{Brookhaven National Laboratory, Upton, NY 11973-5000, U.S.}
\newcommand{\caucr}{University of California - Riverside, Riverside, CA 92521, U.S.}
\newcommand{\ciae}{China Institute of Atomic Energy (CIAE), Beijing, People's Republic of China}
\newcommand{\cns}{Center for Nuclear Study, Graduate School of Science, University of Tokyo, 7-3-1 Hongo, Bunkyo, Tokyo 113-0033, Japan}
\newcommand{\columbia}{Columbia University, New York, NY 10027 and Nevis Laboratories, Irvington, NY 10533, U.S.}
\newcommand{\dapnia}{Dapnia, CEA Saclay, F-91191, Gif-sur-Yvette, France}
\newcommand{\debrecen}{Debrecen University, H-4010 Debrecen, Egyetem t{\'e}r 1, Hungary}
\newcommand{\fsu}{Florida State University, Tallahassee, FL 32306, U.S.}
\newcommand{\gsu}{Georgia State University, Atlanta, GA 30303, U.S.}
\newcommand{\hiroshima}{Hiroshima University, Kagamiyama, Higashi-Hiroshima 739-8526, Japan}
\newcommand{\ihepprot}{IHEP Protvino, State Research Center of Russian Federation, Institute for High Energy Physics, Protvino, 142281, Russia}
\newcommand{\isu}{Iowa State University, Ames, IA 50011, U.S.}
\newcommand{\jinrdubna}{Joint Institute for Nuclear Research, 141980 Dubna, Moscow Region, Russia}
\newcommand{\kaeri}{KAERI, Cyclotron Application Laboratory, Seoul, South Korea}
\newcommand{\kangnung}{Kangnung National University, Kangnung 210-702, South Korea}
\newcommand{\kek}{KEK, High Energy Accelerator Research Organization, Tsukuba, Ibaraki 305-0801, Japan}
\newcommand{\kfki}{KFKI Research Institute for Particle and Nuclear Physics of the Hungarian Academy of Sciences (MTA KFKI RMKI), H-1525 Budapest 114, POBox 49, Budapest, Hungary}
\newcommand{\korea}{Korea University, Seoul, 136-701, Korea}
\newcommand{\kurchatov}{Russian Research Center ``Kurchatov Institute", Moscow, Russia}
\newcommand{\kyoto}{Kyoto University, Kyoto 606-8502, Japan}
\newcommand{\labllr}{Laboratoire Leprince-Ringuet, Ecole Polytechnique, CNRS-IN2P3, Route de Saclay, F-91128, Palaiseau, France}
\newcommand{\lawllnl}{Lawrence Livermore National Laboratory, Livermore, CA 94550, U.S.}
\newcommand{\losalamos}{Los Alamos National Laboratory, Los Alamos, NM 87545, U.S.}
\newcommand{\lpc}{LPC, Universit{\'e} Blaise Pascal, CNRS-IN2P3, Clermont-Fd, 63177 Aubiere Cedex, France}
\newcommand{\lund}{Department of Physics, Lund University, Box 118, SE-221 00 Lund, Sweden}
\newcommand{\muenster}{Institut f\"ur Kernphysik, University of Muenster, D-48149 Muenster, Germany}
\newcommand{\myongji}{Myongji University, Yongin, Kyonggido 449-728, Korea}
\newcommand{\nagasaki}{Nagasaki Institute of Applied Science, Nagasaki-shi, Nagasaki 851-0193, Japan}
\newcommand{\newmex}{University of New Mexico, Albuquerque, NM 87131, U.S.}
\newcommand{\nmsu}{New Mexico State University, Las Cruces, NM 88003, U.S.}
\newcommand{\ornl}{Oak Ridge National Laboratory, Oak Ridge, TN 37831, U.S.}
\newcommand{\orsay}{IPN-Orsay, Universite Paris Sud, CNRS-IN2P3, BP1, F-91406, Orsay, France}
\newcommand{\pnpi}{PNPI, Petersburg Nuclear Physics Institute, Gatchina,  Leningrad region, 188300, Russia}
\newcommand{\riken}{RIKEN, The Institute of Physical and Chemical Research, Wako, Saitama 351-0198, Japan}
\newcommand{\rikjrbrc}{RIKEN BNL Research Center, Brookhaven National Laboratory, Upton, NY 11973-5000, U.S.}
\newcommand{\saispbstu}{Saint Petersburg State Polytechnic University, St. Petersburg, Russia}
\newcommand{\saopaulo}{Universidade de S{\~a}o Paulo, Instituto de F\'{\i}sica, Caixa Postal 66318, S{\~a}o Paulo CEP05315-970, Brazil}
\newcommand{\seoulnat}{System Electronics Laboratory, Seoul National University, Seoul, South Korea}
\newcommand{\stonybrkc}{Chemistry Department, Stony Brook University, SUNY, Stony Brook, NY 11794-3400, U.S.}
\newcommand{\stonycrkp}{Department of Physics and Astronomy, Stony Brook University, SUNY, Stony Brook, NY 11794, U.S.}
\newcommand{\subatech}{SUBATECH (Ecole des Mines de Nantes, CNRS-IN2P3, Universit{\'e} de Nantes) BP 20722 - 44307, Nantes, France}
\newcommand{\tenn}{University of Tennessee, Knoxville, TN 37996, U.S.}
\newcommand{\titech}{Department of Physics, Tokyo Institute of Technology, Oh-okayama, Meguro, Tokyo, 152-8551, Japan}
\newcommand{\tsukuba}{Institute of Physics, University of Tsukuba, Tsukuba, Ibaraki 305, Japan}
\newcommand{\vandy}{Vanderbilt University, Nashville, TN 37235, U.S.}
\newcommand{\waseda}{Waseda University, Advanced Research Institute for Science and Engineering, 17 Kikui-cho, Shinjuku-ku, Tokyo 162-0044, Japan}
\newcommand{\weizmann}{Weizmann Institute, Rehovot 76100, Israel}
\newcommand{\yonsei}{Yonsei University, IPAP, Seoul 120-749, Korea}
\affiliation{\abilene}
\affiliation{\acadsin}
\affiliation{\banaras}
\affiliation{\barc}
\affiliation{\bnl}
\affiliation{\caucr}
\affiliation{\ciae}
\affiliation{\cns}
\affiliation{\columbia}
\affiliation{\dapnia}
\affiliation{\debrecen}
\affiliation{\fsu}
\affiliation{\gsu}
\affiliation{\hiroshima}
\affiliation{\ihepprot}
\affiliation{\isu}
\affiliation{\jinrdubna}
\affiliation{\kaeri}
\affiliation{\kangnung}
\affiliation{\kek}
\affiliation{\kfki}
\affiliation{\korea}
\affiliation{\kurchatov}
\affiliation{\kyoto}
\affiliation{\labllr}
\affiliation{\lawllnl}
\affiliation{\losalamos}
\affiliation{\lpc}
\affiliation{\lund}
\affiliation{\muenster}
\affiliation{\myongji}
\affiliation{\nagasaki}
\affiliation{\newmex}
\affiliation{\nmsu}
\affiliation{\ornl}
\affiliation{\orsay}
\affiliation{\pnpi}
\affiliation{\riken}
\affiliation{\rikjrbrc}
\affiliation{\saispbstu}
\affiliation{\saopaulo}
\affiliation{\seoulnat}
\affiliation{\stonybrkc}
\affiliation{\stonycrkp}
\affiliation{\subatech}
\affiliation{\tenn}
\affiliation{\titech}
\affiliation{\tsukuba}
\affiliation{\vandy}
\affiliation{\waseda}
\affiliation{\weizmann}
\affiliation{\yonsei}
\author{S.S.~Adler}	\affiliation{\bnl}
\author{S.~Afanasiev}	\affiliation{\jinrdubna}
\author{C.~Aidala}	\affiliation{\bnl}
\author{N.N.~Ajitanand}	\affiliation{\stonybrkc}
\author{Y.~Akiba}	\affiliation{\kek} \affiliation{\riken}
\author{J.~Alexander}	\affiliation{\stonybrkc}
\author{R.~Amirikas}	\affiliation{\fsu}
\author{L.~Aphecetche}	\affiliation{\subatech}
\author{S.H.~Aronson}	\affiliation{\bnl}
\author{R.~Averbeck}	\affiliation{\stonycrkp}
% ...
\collaboration{PHENIX Collaboration} \noaffiliation

\date{\today}

\begin{abstract}
The PHENIX experiment has measured mid-rapidity transverse momentum
spectra ($0.4 < p_T < 5.0$~GeV/$c$) of electrons as a function of 
centrality in \auau collisions at \sqrts~=~200~GeV. 
Contributions from photon conversions and from light hadron decays, mainly 
Dalitz decays of $\pi^0$ and $\eta$ mesons, were removed.
The resulting non-photonic electron spectra are primarily due to the 
semi-leptonic decays of hadrons carrying heavy quarks. 
Nuclear modification factors were determined by comparison to non-photonic 
electrons in \pp collisions.
A significant suppression of electrons at high \pt is observed in central 
\auau collisions, indicating substantial energy loss of heavy quarks. 
\end{abstract}


\maketitle

It is well established that neutral pions and charged hadrons are strongly
suppressed at high transverse momentum ($p_T$) in high energy \auau 
collisions \cite{ppg003,ppg014,sup_brahms,sup_phobos,sup_star}. 
The suppression, which is absent in \dau collisions, implies that hard 
scattered partons traversing the medium created in \auau collisions
experience considerable energy loss.
Although high \pt suppression is expected for charm quarks as well, their 
interaction with the medium has been predicted to be smaller than for light 
quarks, {\it i.e.} they should lose a lower fraction of their energy, as their 
large mass decreases the phase space available for gluon radiation, which is 
known as the "dead cone" effect~\cite{dima}. 
If the medium is indeed less opaque to charm quarks they will also participate
less in the collective expansion of the medium, leading to a smaller elliptic 
flow strength $v_2$~\cite{flow} for particles carrying charm quarks compared 
to those solely composed of light quarks.
Such medium effects should be even less pronounced for bottom than for 
charm quarks.

The interaction of heavy quarks with the medium can be studied experimentally 
through systematic measurements of the $p_T$ spectra of open heavy flavor, 
{\it i.e.} hadrons composed of a heavy and a light quark.
While the full reconstruction of $D$ meson decays at the Relativistic Heavy 
Ion Collider (RHIC) is reported for \dau collisions \cite{star_dau},
indirect measurements of open heavy flavor via semi-leptonic decays are 
available for \pp and \dau collisions at 
\sqrts~=~200~GeV~\cite{star_dau,ppg037,phenix_dau} as well as for \auau 
collisions at 130 and 200~GeV~\cite{ppg011,ppg035}.
In \pp collisions, the extracted electron \pt spectrum from heavy flavor decays
is in reasonable agreement with perturbative quantum chromodynamics (pQCD) 
calculations in next-to-leading order. 
However, the data leave room for contributions from further production 
mechanisms in which the heavy quarks are not created in the initial hard 
parton scattering, {\it e.g.} via jet fragmentation~\cite{ppg037}.
In \dau collisions, no indications for strong cold nuclear matter effects
were found~\cite{star_dau,phenix_dau}.
For \auau collisions of different centrality, the total electron yield from 
heavy flavor decays was observed to scale with the nuclear overlap integral
$\langle T_{AA} \rangle$ as expected for point-like pQCD 
processes~\cite{ppg035}.
However, these electrons show an azimuthal anisotropy with respect to the
reaction plane~\cite{ppg040}, consistent with the notion of charm quark flow 
in \auau collisions.
It has been pointed out that if the charm quarks flow along with the bulk of 
the medium, this is evidence for thermalization of charm.
In this situation, the medium modifications of the charm spectrum should
be substantial~\cite{mandt}.

In this Letter, we report on the \pt spectra of non-photonic electrons, 
$(e^+ + e^-)/2$, measured at mid-rapidity ($|\eta| < 0.35$) up to 
$p_T$~=~5~GeV/$c$ by the PHENIX experiment in \auau collisions at 
\sqrts~=~200~GeV. 
The photonic electron background was removed by a {\it cocktail} subtraction, 
in contrast to the {\it converter} subtraction used in~\cite{ppg035}, where 
a subset of the current data sample was analyzed.
The converter method is better suited for a determination of the total 
yield of heavy flavor electrons, while the cocktail subtraction used here 
provides a precision measurement of the spectral shape.
The nuclear modification is then determined by comparing the spectra to those 
in \pp collisions \cite{ppg037}. 

The data used in this analysis were collected by the PHENIX 
detector~\cite{phenix_nim} during the 2001 run of RHIC. 
A coincidence of the beam-beam counters (BBC) and the zero degree calorimeters 
(ZDC) provided the minimum bias trigger ($92.2^{+2.5}_{-3.0}$~\% of the \auau 
inelastic cross section). 
The centrality was determined by the correlation between the multiplicity 
measured by the BBC and the energy of spectator neutrons measured by the ZDC.  
After restricting the vertex range to $|z| < 20$~cm to eliminate background 
originating from the central magnet, a data sample of $25 \times 10^6$ minimum
bias events was analyzed.

For the electron analysis, charged particle tracks were reconstructed with the 
drift chamber and the first layer of pad chambers of the PHENIX east-arm 
spectrometer ($|\eta|<0.35$, $\Delta\phi = \pi/2$), as discussed in detail 
elsewhere~\cite{ppg035}.
Tracks were confirmed by matching hits in the electromagnetic calorimeter 
(EMC) within 2 $\sigma$ in position.
Electron candidates had at least three associated hits in the ring imaging 
\v{C}erenkov detector (RICH).
After an additional cut on the correlation between the momentum $p$ and the 
energy $E$ deposited in the EMC ($-2\sigma < (E-p)/p < 3\sigma$), the only 
background remaining in the electron sample was due to accidental coincidences 
between RICH hits and hadron tracks. 
This background was estimated ($\approx$15~\% at low $p_T$ in central 
collisions, decreasing towards high $p_T$ and for peripheral events) 
and subtracted statistically by an event-mixing method.

\begin{figure}[t]
\includegraphics[width=1.0\linewidth]{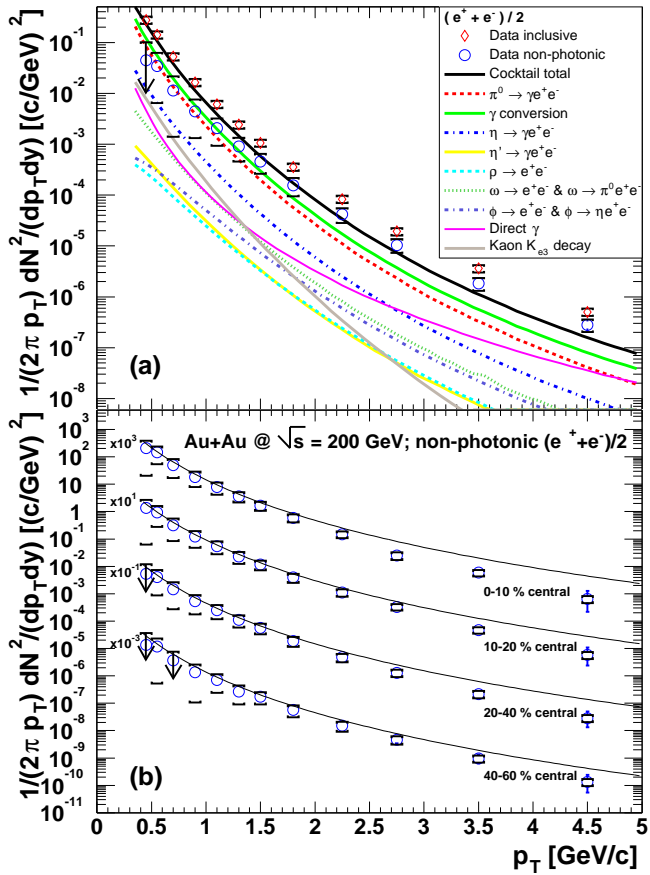}
\caption{\label{fig1}(Color online) Inclusive and non-photonic electron 
invariant yields in minimum bias \auau collisions at \sqrts~=~200~GeV,
compared with contributions from all background electron sources included in 
the cocktail (a). Invariant yields of electrons from heavy flavor
decays for different \auau centrality classes, scaled by powers of ten for 
clarity. Curves are the best fit to the \pp reference scaled with the 
appropriate nuclear overlap integrals $\langle T_{AA} \rangle$ (b). 
The error bars (brackets) correspond to statistical (systematic) uncertainties 
in both panels.}
\end{figure}

The raw electron spectra were corrected as a function of $p_T$ for geometrical 
acceptance and reconstruction efficiency~\cite{ppg035}.
The multiplicity dependent efficiency loss was estimated by embedding simulated
electrons into real events.
This loss does not depend on \pt and increases from 5 to 26~\% from peripheral 
to central collisions. 
The $1\sigma$ systematic uncertainty of all corrections is 11.8~\%, after
correction for the effect of finite bin width in $p_T$. 
The fully corrected inclusive electron spectrum is shown in Fig.~\ref{fig1}(a)
for minimum bias collisions.

The spectra of electrons from heavy flavor decays were determined by 
subtracting cocktails of background contributions from other sources from 
the inclusive data.
The most important background is the $\pi^0$ Dalitz decay which was 
calculated individually for each centrality class with a hadron decay 
generator using parameterizations of measured $\pi^0$~\cite{ppg014} and 
$\pi^\pm$~\cite{ppg026} spectra as input. 
The spectral shapes of other light hadrons $h$ were obtained from the pion 
spectra, assuming a universal spectrum in $m_T = \sqrt{p_T^2 + m_h^2}$.
Within this approach the ratios $h/\pi^0$ are constant at high $p_T$ with 
the values~\cite{ppg011}: 
$\eta/\pi^0 = 0.45 \pm 0.10$, $\rho/\pi^0 = 1.0 \pm 0.3$, 
$\omega/\pi^0 = 1.0 \pm 0.3$, $\eta'/\pi^0 = 0.25 \pm 0.08$, and 
$\phi/\pi^0 = 0.40 \pm 0.12$. 
Only the $\eta$ contribution is of any practical relevance, and the chosen
parameterization is in good agreement with the measured $\eta$ meson 
spectra~\cite{ppg051}.
Another major electron source is the conversion of photons, mainly from 
$\pi^0 \rightarrow \gamma\gamma$ decays, in material in the acceptance
($\approx$1\% $X/X_0$). 
The spectra of electrons from conversions and Dalitz decays are very 
similar.
In a GEANT simulation of $\pi^0$ decays, the ratio of conversion electrons to 
Dalitz electrons was determined to be $1.25 \pm 0.10$, essentially $p_T$ 
independent. 
Contributions from photon conversions from other sources were taken into 
account as well. 
Electrons from kaon decays ($K_{e3}$), determined in a GEANT simulation based 
on measured kaon spectra~\cite{ppg026}, and electrons from external as well as 
internal conversions of direct photons~\cite{vogelsang,ppg042} were included.

All background sources are compared with the inclusive data in 
Fig.~\ref{fig1}(a). 
Further background from $J/\psi \rightarrow e^+e^-$ decays and 
from Drell-Yan pairs \cite{drellyan} is negligible.
A possible low mass dilepton enhancement through 
$\pi+\pi \rightarrow \rho \rightarrow e^+e^-$, as reported in Pb+Pb
collisions at the SPS~\cite{ceres}, would constitute another background
source which is neglected here since the estimated $\rho$ contribution in the 
absence of enhancement is small ($< 1 \%$ at all $p_T$).
The total cocktail systematic uncertainty increases from 10~\% (at 
\pt~=~0.4~\gevc) to 15~\% (at \pt~=~5~GeV/$c$), dominated by the systematic 
error of the pion input spectra ($\approx$ 8-10~\%). 
Other systematic uncertainties, mainly the $\eta/\pi^0$ normalization and, 
at high $p_T$, the contribution from direct radiation, are much smaller.
The background cocktail calculated here and the photonic electron background 
measured via the converter method~\cite{ppg035} agree within 10~\%.

After subtracting the cocktail from the inclusive electron data, the invariant
spectrum of electrons from heavy flavor decays is shown in Fig.~\ref{fig1}(a)
for minimum bias collisions.
For $p_T > 2$~\gevc the signal to background ratio is larger than one.
Fig.~\ref{fig1}(b) shows the electron spectra from heavy flavor decays in 
four centrality classes, 0-10~\%, 10-20~\%, 20-40~\%, and 40-60~\% central 
collisions. 
More peripheral collisions have insufficient electron statistics to reach 
$p_T$~=~5~GeV/$c$.

PHENIX has also measured electrons from heavy flavor decays in \pp collisions
at $\sqrt{s} = 200$~GeV~\cite{ppg037}.  
The curves shown in Fig.~\ref{fig1}(b) depict the best fit of the 
corresponding spectrum from \pp collisions, scaled by the nuclear overlap
integral $\langle T_{AA} \rangle$ calculated within a Glauber 
model~\cite{ppg014} for each \auau centrality class.
At low $p_T$ the \auau spectra are in reasonable agreement with the \pp fit 
in all centrality bins, but a clear suppression of the spectra in \auau with 
respect to \pp develops towards high $p_T$.

To quantify this effect we calculate for each individual bin in \pt the 
nuclear modification factor \raa defined as
\begin{equation}
R_{AA} = \frac{dN_{Au+Au}}{\langle T_{AA} \rangle \times d\sigma_{p+p}}
\end{equation}
where $dN_{Au+Au}$ is the differential electron yield from heavy flavor decays 
in \auau collisions and $d\sigma_{p+p}$ is the corresponding differential cross
section in \pp collisions~\cite{ppg037} in any given \pt bin.

\begin{figure}[t]
\includegraphics[width=1.0\linewidth]{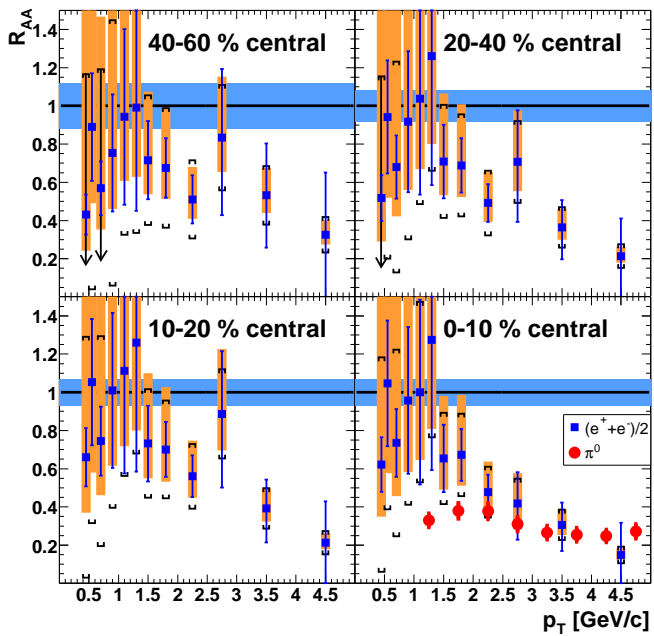}
\caption {\label{fig2}Nuclear modification factor \raa for electrons from 
heavy flavor decays as function of \pt in \auau collisions at \sqrts~=~200~GeV
for the different centrality classes. The error bars are statistical only. 
Error brackets (boxes) indicate the systematic errors related to the 
uncertainties in the \auau (p+p) measurements. The bands around one show the 
relative systematic uncertainties in $T_{AA}$. For the most central collisions 
the $\pi^0$ \raa is shown for comparison~\cite{ppg014}. For these data, a 
13~\% $p_T$ independent systematic uncertainty (not plotted) represents the 
uncertainty in $\langle T_{AA} \rangle$ and in the $\pi^0$ yield 
normalization.}
\end{figure}

Fig.~\ref{fig2} shows \raa as a function of \pt in the four \auau centrality 
classes.
At low $p_T$, the electron \raa is consistent with one within substantial
uncertainties in all centrality classes, in agreement with the observation 
of binary collision scaling of the total charm yield in \auau collisions at 
RHIC~\cite{ppg035}.
Since the ratio of electrons from heavy flavor decays to background 
increases with increasing $p_T$, the systematic uncertainties of \raa
decrease towards high $p_T$.
\raa falls well below one for electron $p_T \ge 2$~GeV/$c$, providing clear
evidence for heavy quark medium modifications.
The observed high \pt suppression is most significant for central collisions.
However, the limited statistics do not allow to quantify the centrality 
dependence of heavy quark medium modifications.
At the highest $p_T$, the electron \raa becomes as small as that for 
$\pi^0$~\cite{ppg014}, indicating substantial energy loss of heavy quarks in 
the medium.
It is important to note that electrons at a given \pt originate from decays 
of higher \pt $D$ or $B$ mesons, making model independent comparisons of \raa 
for light and heavy quarks impossible.

The observed \raa is remarkable, as electrons with $p_T > 3.5$~\gevc are
expected to include significant contributions from $B$ meson decays, and $B$ 
mesons should suffer less than $D$ mesons from medium modifications.
Depending on their time scales, mechanisms by which heavy quarks are produced
after the initial hard parton scattering, such as gluon splitting in jets,
might lead to an attenuation at high \pt which then is due to a mixture of 
light parton and heavy quark energy loss in the medium created at RHIC.

\begin{figure}[t]
\includegraphics[width=1.0\linewidth]{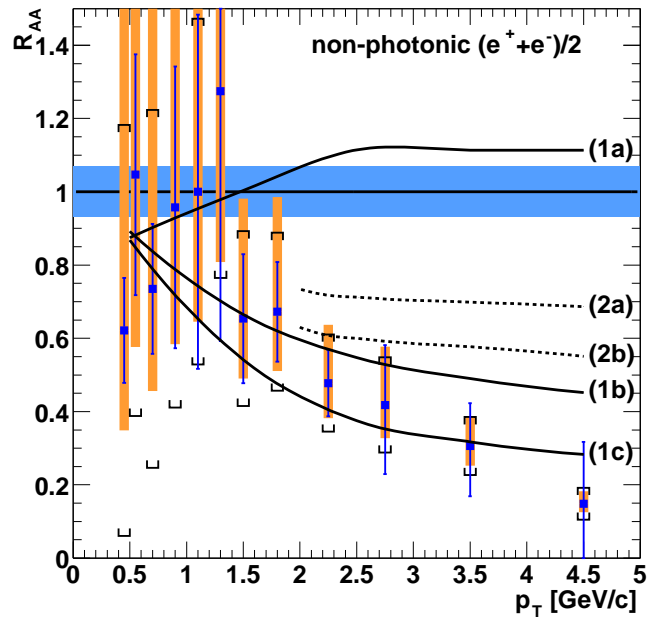}
\caption{\label{fig3}Nuclear modification factor \raa for electrons from 
heavy quark decays as function of \pt for the 10~\% most central \auau
collisions at \sqrts~=~200~GeV in comparison with predictions from models 
incorporating charm quark energy loss. The curves (1a-c) and (2a-b) are taken
from~\cite{armesto} and \cite{magdalena}, respectively, where contributions 
from $B$ meson decays are included in (2a-b) only. Experimental uncertainties 
are shown as described in Fig.~\ref{fig2}.}
\end{figure}

Fig.~\ref{fig3} confronts current model calculations \cite{armesto,magdalena} 
utilizing induced gluon radiation as the heavy quark energy loss mechanism
with the data for the 10~\% most central collisions.
The three curves (1a-c) include electrons from charm decays 
only~\cite{armesto}.
They correspond to different values of the time-averaged transport coefficient 
$\hat{q}$, which denotes the average squared transverse momentum transferred 
from a hard parton per unit path length while traversing the medium and, as 
such, is proportional to the density of scattering centers in the medium.
Curve (1a) applies for the case without the presence of any medium causing
heavy quark energy loss ($\hat{q}$~=~0~GeV$^2$/fm).
The $\hat{q}$ values of 4 and 14~GeV$^2$/fm, which correspond to the curves
(1b) and (1c), lead to light quark energy losses which bracket the observed
high $p_T$ suppression of neutral pions and charged hadrons. 
Predictions for charm energy loss from~\cite{armesto} for medium densities 
at the extreme high end of those allowed by the observed light quark energy 
loss are consistent with the electron data.
Contributions from bottom decays, which are expected to be significant for 
$p_T > 3$~GeV/$c$, should lead to an increase of the predicted \raa since
$b$ quarks are presumably less affected by energy loss than $c$ 
quarks~\cite{dima}.
The curves (2a-b) are taken from~\cite{magdalena}. 
They include electrons from both $D$ and $B$ meson decays and correspond to 
initial gluon densities of $dN_g/dy = 1000 (3500)$ for curve (2a(b)), 
respectively, which again lead to light parton energy losses bracketing the 
observed high \pt pion suppression.
However, at high \pt the predicted \raa for electrons from heavy flavor decays 
is larger than observed.
The present data pose a challenge to existing calculations of radiative
energy loss in the medium produced at RHIC, and will help to distinguish
between different energy loss scenarios.

In conclusion, we have measured electron spectra from heavy flavor 
decays in \auau collisions at \sqrts~=~200~GeV.
In central collisions, nuclear modification factors $R_{AA} << 1$ are observed
at high $p_T$, providing clear evidence for strong medium effects.
Current models involving energy loss via induced gluon radiation for heavy
quarks traversing the medium created in heavy ion collisions at RHIC are 
challenged by the data even considering extremely high medium densities.

%\section{Acknowledgements}   % Run-2 short form for PRL
We thank the staff of the Collider-Accelerator and Physics
Departments at BNL for their vital contributions.  We acknowledge
support from the Department of Energy and NSF (U.S.A.), 
MEXT and JSPS (Japan), CNPq and FAPESP (Brazil), NSFC (China), 
CNRS-IN2P3 and CEA (France), 
BMBF, DAAD, and AvH (Germany), 
OTKA (Hungary), DAE and DST (India), ISF (Israel), 
KRF and CHEP (Korea), RMIST, RAS, and RMAE (Russia), 
VR and KAW (Sweden), U.S. CRDF for the FSU, 
US-Hungarian NSF-OTKA-MTA, and US-Israel BSF.

\def\IJMPA{{Int. J. Mod. Phys.}~{\bf A}}
\def\JPG{{J. Phys}~{\bf G}}
\def\NCA{Nuovo Cimento}
\def\NIM{Nucl. Instrum. Methods}
\def\NIMA{{Nucl. Instrum. Methods}~{\bf A}}
\def\NPA{{Nucl. Phys.}~{\bf A}}
\def\NPB{{Nucl. Phys.}~{\bf B}}
\def\PLB{Phys. Lett. B}
\def\PLC{Phys. Repts.\ }
\def\PRL{Phys. Rev. Lett.\ }
\def\PRD{Phys. Rev. D}
\def\PRC{Phys. Rev. C}
\def\ZPC{{Z. Phys.}~{\bf C}}
\def\etal{{\it et al.}}

\end{document}